# Ultrafast adsorbate excitation probed with sub-ps resolution XAS

Elias Diesen,*,[1] Hsin-Yi Wang,[2] Simon Schreck,[2] Matthew Weston,[2] Hirohito Ogasawara,[3] Jerry LaRue,[4] Fivos Perakis,[2] Martina Dell'Angela,[5] Flavio Capotondi,[6] Luca Giannessi,[6] Emanuele Pedersoli,[6] Denys Naumenko,[6] Ivaylo Nikolov,[6] Lorenzo Raimondi,[6] Carlo Spezzani,[6] Martin Beye,[7] Filippo Cavalca,[2] Boyang Liu,[2] Jörgen Gladh,[2,3] Sergey Koroidov,[2] Piter S. Miedema,[7] Roberto Costantini,[5,8] Tony F. Heinz,[3,9] Frank Abild-Pedersen,[1] Johannes Voss,[1] Alan C. Luntz,[1] Anders Nilsson[2]

[1]*SUNCAT Center for Interface Science and Catalysis, SLAC National Accelerator Laboratory, 2575 Sand Hill Road, Menlo Park, California 94025*

[2]*Department of Physics, AlbaNova University Center, Stockholm University, SE-10691 Stockholm, Sweden*

[3]*SLAC National Accelerator Laboratory, 2575 Sand Hill Road, Menlo Park, California 94025*

[4]*Schmid College of Science and Technology, Chapman University, Orange, California 92866*

[5]*CNR-IOM, SS 14 – km 163.5, 34149 Basovizza, Trieste, Italy*

[6]*FERMI, Elettra-Sincrotrone Trieste, SS 14 – km 163.5, 34149 Basovizza, Trieste, Italy*

[7]*Deutsches Elektronen-Synchrotron DESY, Notkestrasse 85, 22607 Hamburg, Germany*

[8]*Physics Department, University of Trieste, Via Valerio 2, 34127 Trieste, Italy*

[9]*Department of Applied Physics, Stanford University, Stanford, California 94305*

* E-mail: ediesen@stanford.edu

**ABSTRACT**

We use a pump-probe scheme to measure the time evolution of the C K-edge X-ray absorption spectrum (XAS) from CO/Ru(0001) after excitation by an ultrashort high-intensity optical laser pulse. Due to the short duration of the X-ray probe pulse and precise control of the pulse delay, the excitation-induced dynamics during the first ps after the pump can be resolved with unprecedented time resolution. By comparing with theoretical (DFT) spectrum calculations we find high excitation of the internal stretch and frustrated rotation modes occurring within 200 fs of laser excitation, as well as thermalization of the system in the ps regime. The ~100 fs initial excitation of these CO vibrational modes is not readily rationalized by traditional theories of nonadiabatic coupling of adsorbates to metal surfaces, *e. g.* electronic frictions based on first order electron-phonon coupling or transient population of adsorbate resonances. We suggest that coupling of the adsorbate to non-thermalized electron-hole pairs is responsible for the ultrafast initial excitation of the modes.

**MAIN TEXT**

The vibrational, rotational, and translational motions of adsorbates are the fundamental physical processes behind chemical reactions on surfaces. Understanding the details of adsorbate dynamics can therefore be essential for describing heterogeneous catalytic reactions, with the eventual goal of tailoring catalysts for high efficiency and selectivity. In particular, how energy is transferred from the bulk substrate to the adsorbate modes is key to understanding selectivity in catalysis. Since surface chemical reactions proceed through several elementary dynamical steps on different time scales in parallel with competing energy relaxation processes, studying these in a time-resolved manner can give knowledge about the elementary adsorbate-surface interaction and energy transfer processes.

Exciting an adsorbate/substrate system with an intense femtosecond optical pulse and then probing the dynamical evolution of the adsorbate by time-resolved optical techniques, such as sum-frequency generation (SFG) [1–6] and two-pulse correlation (2PC) [7–9] has a modest history. More recently, X-ray absorption and emission spectroscopy (XAS/XES) [10–14], has also become a powerful approach for gaining insight into the transient excitation of adsorbate vibrations and the consequent diffusion, desorption, and final chemical reactions. X-ray spectroscopies are inherently element-specific local probes and thus particularly suited for studying any process that affects the electronic structure and chemical environment of a specific atom [15]. In particular, an X-ray free-electron laser (XFEL) delivering femtosecond X-ray pulses allows probing dynamical changes of the adsorbate electronic structure induced by the femtosecond optical laser. Activation of atomic oxygen [12] and carbon monoxide [10], as well as driving of CO oxidation [13], are examples of reactions that have been studied with this technique.

In this work, we report experimental C K-edge XAS of CO/Ru(0001) during and after pulsed optical laser excitation, probed with a time resolution of ~100 fs in a conventional pump-probe setup. This is possible due to the ultrashort duration of XFEL pulses (at DIPROI beamline of the FERMI facility in Elettra, Italy). The XFEL is seeded by a tunable solid state laser system and the photon energy is multiplied through a two-stage high-gain harmonic generation (HGHG) process [16]. The polarized soft X-ray pulses with less than 50 fs duration are delivered to the sample at a repetition rate of 10 Hz as the "probe". Meanwhile, a portion of the seed laser is separated and coupled into the experimental chamber collinearly with the X-rays, acting as the "pump" with the same repetition rate. Since the seed and the pump originate from the same laser pulse, the time jitter between the pump laser and the XFEL X-ray pulse is less than 6 fs [17]. The aforementioned properties allow direct detection of the adsorbate electron structure as influenced by internal excitations before the substrate phonons have been excited, which takes around 1 ps. Since CO chemisorbs on Ru(0001) in an upright geometry at the on-top site (with the C atom coordinated toward the Ru atom) [18], the dipole selection rule gives the strongest moment for the C K-edge 1s → 2π* transition when the electric field vector (E-vector) is parallel to the surface. Using linearly polarized X-rays and measuring with the E-vector oriented either in-plane (lin H) or out-of-plane (lin V) with respect to the metal surface, we can thus monitor how the electronic structure and molecular orientation temporally changes following the optical laser pumping. The optical laser is p-polarized, precluding any direct optical excitation of the 2π*. The C K-edge, due to similar intramolecular potential energy curves in the ground and core-excited states, gives only small Franck Condon factors for C-O vibrational excitation in the x-ray transition [19,20], and thereby provides excellent spectral contrast in comparison to the much broader O K-edge used in most previous work [10,11,21].

Recently, CO desorption from CO/Ru was studied on ps time scales, and a weakly adsorbed precursor state [11,22] was directly identified and its evolution described [14]. In this paper, we focus on the ultrafast dynamics occurring in the CO within the very first picosecond following optical laser excitation of the Ru. The high spectral resolution and controlled pump-probe delay, together with detailed spectrum simulations, allow us to identify the excitation of specific adsorbate modes and their initial excitation times. We find significantly faster energy transfer to the adsorbate than conventional non-adiabatic theoretical approaches suggest, and we postulate that non-thermal electron-hole (e-h) pairs initially created by the optical laser may account for the ultrafast initial excitations of these modes.

The Ru(0001) crystal was first cleaned by standard procedures [14], and a saturated CO adlayer (~0.66 ML) on Ru was prepared by pressurizing the UHV chamber with $1.0 \times 10^{-8}$ torr of CO while cooling the crystal down to liquid nitrogen temperature [23]. The crystal sample was pumped using 400 nm optical laser pulses (fluence of ~91 J/m$^2$) at a grazing incidence angle of around 2° with duration of 100 fs. A fresh sample spot for each shot was realized by continuously scanning the sample. A CO background pressure of $1.0 \times 10^{-8}$ torr was used to ensure a fully recovered surface adlayer once the same spot was revisited after scanning the complete sample. XAS was recorded by detecting X-ray fluorescence, using a previously described detector setup [24], and the incident energy was stepwise changed across the carbon 1s → 2π* resonance edge in the range of 285–290 eV. The XAS was thus plotted by averaging the total number of emitted photons recorded as a function of the incident energy for every desired pump–probe delay. The definition of time zero delay is discussed in the supplementary information (Suppl. A [25]). Further details of the experimental method can be found elsewhere [14].

The top panel of Fig. 1 shows the lin H XAS spectra evolving with pump-probe delay time. The optical unpumped spectrum (averaged between -300 – 0 fs) depicts a strong C 1s → 2π* resonance peak centered at around 288 eV. For delays 0 – 100 fs following the laser pump pulse, a sudden decrease of intensity at the peak center and an increase of intensity on the lower energy side of the spectrum is clearly seen, while the main peak energy stays the same. However, in the next 100 fs delay a redshift of about 80 meV of the peak center can be identified. No further rapid changes are found in the following delays, indicating a slower energy equilibration after the initial excitation. The lin V spectra (Fig. 1(b)), on the other hand, indicates that the aforementioned observation in the lin H spectrum at 0-100 fs delay cannot be ascribed to the molecule tilting since no corresponding intensity increasing at 288 eV can be seen in the lin V spectrum. Instead, a mild increase of overall intensity in lin V spectra can be seen in the following delays, up to about 1 ps.

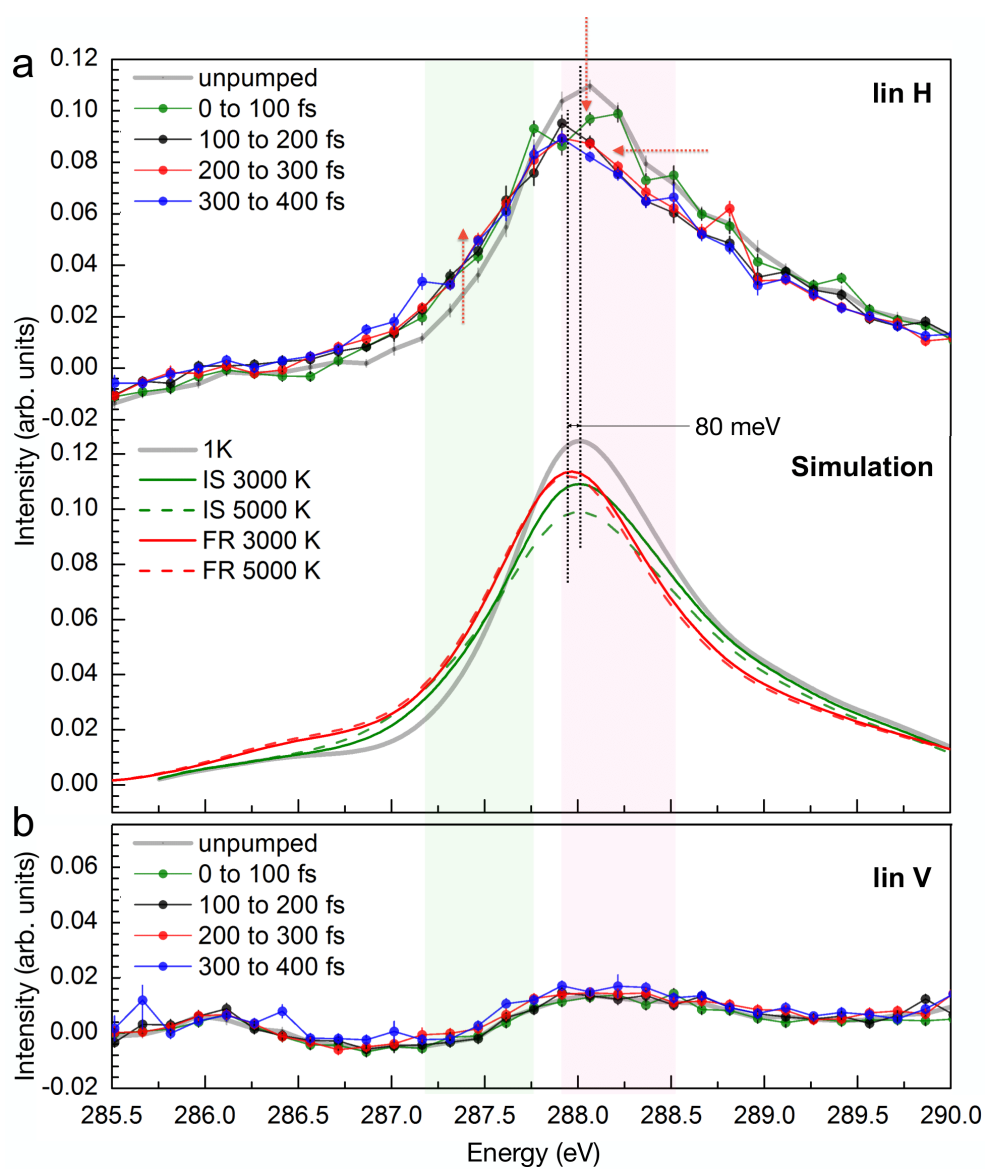

FIG. 1. (a) Measured lin H spectra, averaged over the indicated pump-probe delay ranges, together with calculated spectra with excited internal stretch (IS) and frustrated rotation (FR) modes. (b) Measured lin V spectra. (The negative intensity is due to background subtraction.)

To study the time evolution in detail, the intensities are integrated over two energy regions (main peak region in the pink area and lower energy region in the green area) chosen in order to give a clear view of the time-dependence. The time evolution of these integrated intensities is shown in Fig. 2. In addition, the lin H and lin V spectra at longer time delays are shown in Suppl. B [25]. We first focus on the main peak region. Fig. 2(a) shows a sudden decrease of intensity in the first 0-100 fs and 100-200 fs delays of the lin H spectrum due to the main peak intensity dropping and peak shifting, respectively (cf. Fig. 1(a)), followed by a gradually decreasing intensity after 200 fs and a corresponding increasing intensity in the lin V spectrum with time. On the lower energy side (Fig. 2(b)), a sudden increase of intensity in the first 0-100 fs delay followed by a gentle increase of intensity after 100 fs can be observed in the lin H spectra, while for lin V spectra no obvious changes in 0-100 fs delay can be seen, yet a similar intensity increasing trend as in the lin H spectra shows up after 100 fs. We can thus distinguish two processes occurring on different time scales, one very rapid which is completed after around 200 fs and mainly affects the lin H spectrum, and one slower affecting both polarizations.

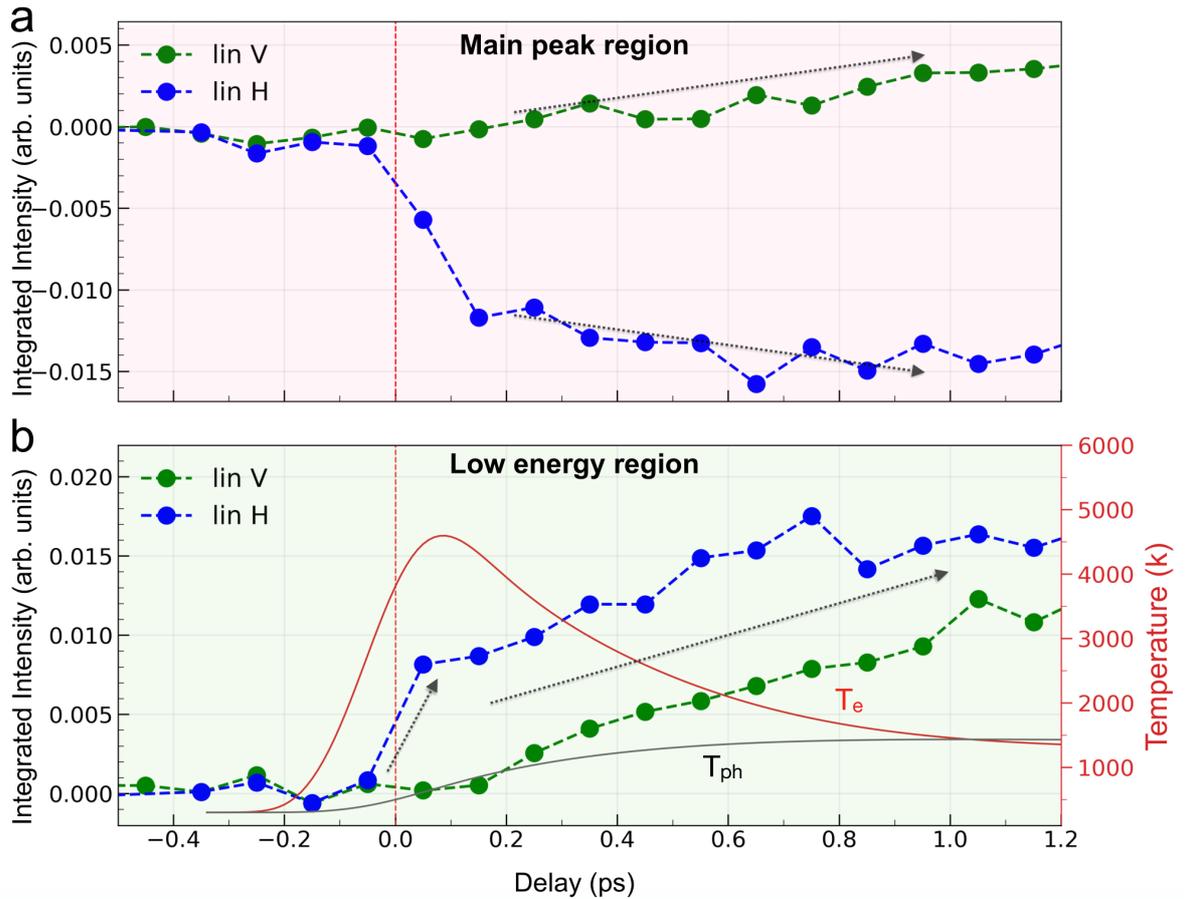

FIG. 2. Time evolution of the integrated spectral intensity, minus the unpumped intensity, in (a) the main peak region and (b) low energy region, respectively (cf. Fig. 1). Also shown in (b) are the electron ($T_e$) and phonon ($T_{ph}$) temperatures from the two-temperature model.

To interpret the time evolution of the spectra, we perform density functional theory (DFT) calculations using the Quantum Espresso [26,27] code, with spectra calculated using the xspectra [28–30] code. The general method for spectral calculation and treatment of the core-hole has been described elsewhere [31]. A 4×4×4 periodic slab was used for all spectrum calculations with a single CO molecule adsorbed at an on-top site, corresponding to ¼ ML coverage instead of the experimental 0.66 ML. However, we have checked that our spectroscopy simulations are not sensitive to the absolute coverage (see Suppl. F [25]). For further computational details, see Suppl. C [25].

Previous theoretical studies of the first order electron-phonon coupling of CO adsorbed on transition metal surfaces, *e g.* CO/Cu(100), CO/Ni(111) and CO/Pt(111), suggested that the internal C-O stretch (IS) and the frustrated CO rotation (FR) modes are the modes most likely to be significantly excited directly by thermal metal electron-hole pair excitations [32]. Our theoretical studies of the first order electron-phonon coupling for CO/Ru (see Suppl. K [25]) confirm this pattern, with the IS exhibiting the strongest coupling followed by that for FR. To calculate the XAS for internal stretch vibrationally excited states of the CO molecule at the surface, we determine the potential energy curve for the electronic ground state by varying the C-O distance and calculating the DFT total energy for each point. We then fit a Morse potential to the data, which gives the eigenstates and an analytical representation of the vibrational eigenfunctions. The potential for the internal stretch mode with core-excited C was determined in the same way, with the C pseudopotential containing an explicit core-hole. The unit cell was kept neutral by adding an electron at the Fermi level, since the metal electrons screen the core-hole very rapidly [33]. The Franck-Condon factors for the C 1s → $2\pi^*$ transition can then be calculated analytically. We calculate the spectra at several displacements and average them together with the appropriate weight for the vibrational eigenstate to get the spectrum for each vibrational state (for details, see Suppl. D and E [25]). The frustrated rotation mode was treated similarly, but with a harmonic potential assumed for the description of the mode. Finally, the spectra for the different vibrational states were added together, with weights given by the Boltzmann distribution for a desired temperature.

The calculated spectra for different vibrational modes at various thermal excitations are shown in Fig. 1. We first note that the agreement of the ground state theoretical spectrum with the experimental spectrum at negative delay is excellent (see also Fig. S4), giving confidence to the spectrum simulations. The calculated spectra for elevated temperatures of the internal stretch (IS) mode show a broadening of the peak and decrease of the maximum intensity, however the peak position remains unchanged. High temperature in the frustrated rotation (FR) mode however gives an 80 meV redshift, in excellent quantitative agreement with the experimental results for ~200 fs delay. We therefore identify the IS excitation as inducing the initial decrease of the peak and the FR excitation as causing the redshift of the peak. It might be expected that FR motion would also give a strong signal in the vertical polarization direction, since tilting the molecule means the C 1s → 2π* dipole moment will have nonzero projection in the vertical direction. However as seen also in Fig. S5 [25] the actual response is quite small. During pure FR excitation the molecular center-of-mass remains fixed, and there is no simple direct relation between the two polarization signals (Fig. S5). In addition, the slow increase we see in the vertical signal over the first ps is consistent with the anticipated complicated dynamics en route to overall thermalization of the system. For comparison, the results of the commonly used two-temperature model is also shown in Fig. 2(b) (see also Suppl. G [25]). It is immediately clear that the initial response in the lin H signal (blue dotted line) is faster than the rise of the phonon temperature (black line), and thus must be electron-driven, while the lin V signal (green dots) increase is due to more complicated dynamics involving both electrons and phonons (see discussion later). After ~ 1 ps, many phonons are excited and the surface geometry becomes much less well-defined [3]. There is therefore a large number of adsorption sites and rotational states available, giving a slow increase of resonance intensity projected on the vertical direction (lin V) and a complementarily decreasing intensity in lin H spectra in the main peak region at 0.2 - 1 ps delay (Fig. 2(a)). The spectral changes at 1 ps can be qualitatively reproduced by simulating spectra from *ab initio* molecular dynamics at a temperature of 1300 K (see Suppl. H [25]), suggesting that near-equilibration between all modes occurs by ~1 ps. Desorption starts to be visible in the spectrum after a few ps [14], beyond the time range studied in this paper.

The mechanism behind the very fast adsorbate initial IS and FR excitation, which occurs within 200 fs, is unclear. First order electron-phonon coupling calculations of the frictional damping of thermalized metal electron-hole pairs suggest vibrational lifetimes τ ~ 2 - 4 ps for the IS and FR modes for CO/Ru (see Suppl. K [25]). A simple one dimensional classical master equation approach indicates that a coupling strength of ≥30 $cm^{-1}$ (corresponding to τ~200 fs) would be needed to obtain an energy transfer fast enough to describe the observed initial excitations in these modes (Suppl. I [25]). We suggest that the most likely reason for the rapid initial excitation is that the highly excited non-thermal e-h pair distribution, initially created in the substrate by laser absorption, could give a stronger coupling to adsorbate motion than friction models based on first order electron-phonon coupling where a thermalized e-h pair bath is assumed to excite the adsorbate. Possible evidence for the contribution of non-thermalized e-h pair to adsorbate excitation has previously been suggested for oxygen activation in a O/CO/Ru system [34], where the non-thermalized electron population right after optical excitation coupled more efficiently to an unoccupied adsorbate resonance (LUMO) compared to a thermalized hot Fermi–Dirac distribution, which does not necessarily form within the first 100 fs [35].

The LUMO $2\pi^*$ resonance for CO/Ru is found around 5 eV above the Fermi level [36], while the optical laser energy is only 3.1 eV (i.e., 400 nm). Even though the $2\pi^*$ resonance is broad and a remnant of its resonance could extend to 3.1 eV, significant direct population of this resonance - as in ion-resonance models responsible for Desorption Induced by (Multiple) Electronic Transitions [37] (DIET/DIMET) - is highly unlikely to give efficient energy transfer. Even if we assumed that the adsorbate LUMO $2\pi^*$ resonance could be populated (e.g., by photo-excitation of initially created hot electrons in the laser pulse) we find that a simple dynamical model gives only little energy transfer into the motion of the adsorbate (Suppl. J [25]), thus making resonant population of the adsorbate LUMO an inefficient excitation mechanism for CO/Ru at 400 nm. However, the non-thermal e-h pair distribution could still excite adsorbate modes more efficiently than thermal e-h pairs, since e-h pair states far from the Fermi level may couple more strongly than close to it.

In this paper, we only discuss the anomalous fast initial excitations of the IS and FR modes observed in the C XAS spectra compared to those predicted by first order electron-phonon coupling, and the description after full thermalization. We make no attempt to describe the route between the two as that is driven by complicated coupled electron and phonon dynamics. Theoretical studies of the temperature dependence of the CO/Cu(100) IS lineshape and its evolution following fs optical laser excitation even imply that first order electron-phonon coupling is not sufficient to describe the temporal dynamics of the coherently excited $v = 1$ state of IS [38,39]. Extending the electron-phonon coupling to second order opens up an entirely new important class of energy transfer processes, so-called electron-mediated phonon-phonon couplings (EMPPC). While we have no definitive argument that these EMPPC processes are as important for CO/Ru as for CO/Cu(100), we believe that they likely are so that a complete understanding of the dynamics occurring between initial excitation and the fully thermalized system will ultimately require a multidimensional dynamical model including all of the EMPPC and direct phonon couplings. However, it is unlikely that their addition can account for the ultrafast initial mode excitations responsible for the changes in the XAS. For CO/Cu(100), the fastest decay processes for the wavevector $q \sim 0$, $v = 1$ state of IS are due to the first order electron-phonon coupling and dephasing of the $q \sim 0$ into other $q$ states of the same IS vibration (denoted as IS $\rightarrow$ IS) [39]. However, unlike photon excitation of an adsorbate vibrational mode, e-h pair scattering excites the adsorbate modes with all $q$. In addition, since XAS is a local probe at the C, the XAS spectrum is independent of the phase of vibrations contributing to the lineshape so that the IS $\rightarrow$ IS process does not contribute to the dynamics observed. Therefore, if EMPPC in CO/Ru are similar to those in CO/Cu, inclusion of EMPPC does not rationalize the ultrafast initial changes to the XAS compared to those predicted by first order electron-phonon coupling. Since optically created electron-hole pairs take $\sim$ 100 fs to thermalize in Ru [35], we instead suggest that non-thermalized electron-hole pairs within the first $\sim$100 fs have a stronger coupling to the adsorbate and preferentially excite the IS and FR modes within this time scale.

We summarize our main observations in Fig. 3. We can tentatively identify an initial excitation of the IS within 100 fs of the laser pulse, followed by excitation of the FR mode during the subsequent 100 fs. Thermalization of the substrate phonons, and heating of the adsorbate translational modes, proceeds over the first ps, consistent with the commonly used two-temperature model. The initial excitations are significantly faster than what can be accounted for in non-adiabatic friction models with theoretically obtained friction coefficients, pointing to a lack of current theoretical descriptions of the coupling between highly excited electron distributions and adsorbate dynamics. Our observation is however in line with what has been inferred previously from experimental SFG [2,3,5,6] and 2PC [9] studies. A detailed explaination poses a challenge for future theoretical and experimental work.

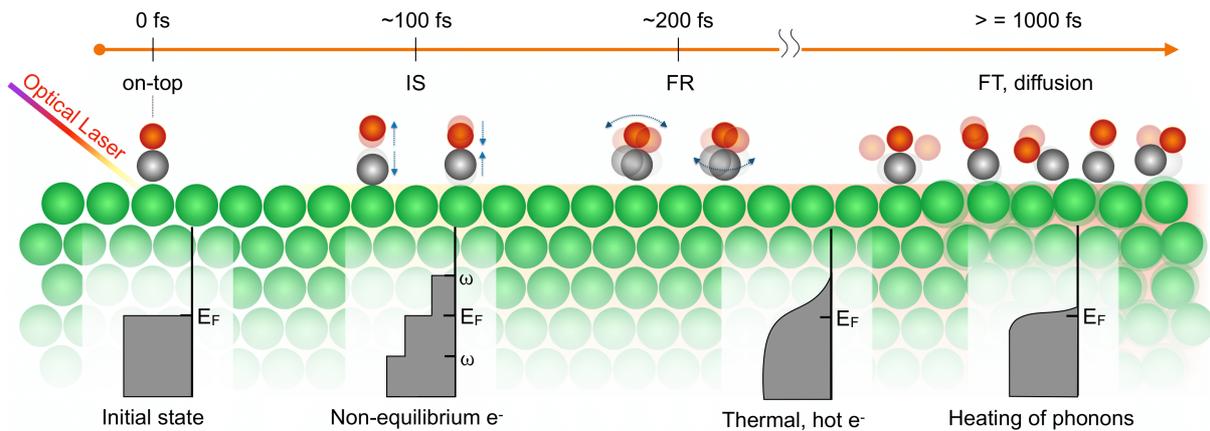

FIG. 3. Sketch of the system evolution during the first picosecond. Atom colors: O red, C grey, Ru green.

This study shows the possibilities of the sub-ps time resolution allowed by the well-controlled pump-probe delay in a seeded FEL setup. Compared to the 2PC technique, where sub-ps dynamics can only be inferred indirectly by its influence on subsequent dynamics, and to SFG, where one can only probe a specific degree of freedom (typically the IS mode at $q \sim 0$), this technique gives uniquely detailed information about the immediate response of the adsorbate local electronic structure to the driving pulse. We have thus been able to directly measure the evolution with time as it occurs, and together with DFT spectral calculations this allows a detailed analysis of the initial excitation and fully thermalized state. We expect that similar methods could be used to investigate the questions raised in this work regarding the role of non-thermal electrons and the seemingly strong frictional coupling.

This research was supported by the Knut and Alice Wallenberg Foundation under Grant No 2016.0042, the Swedish Research Council under Grant No 2013-8823, U.S. Department of Energy, Office of Science, Office of Basic Energy Sciences, Chemical Sciences, Geosciences, and Biosciences Division, Catalysis Science Program to the Ultrafast Catalysis FWP 100435 at SLAC National Accelerator Laboratory under Contract DE-AC02-76SF00515. This research used resources of the National Energy Research Scientific Computing Center, a DOE Office of Science User Facility supported by the Office of Science of the U.S. Department of Energy under Contract DE-AC02-05CH11231. The authors acknowledge the continuous support of FERMI team during the setting and the operation of the FEL source for the experiment. M.B and P.S.M acknowledge funding from the Helmholtz association (VH-NG-1005). M.D.A and R.C. acknowledge support from the SIR grant SUNDYN [Nr RBSI14G7TL, CUP B82I15000910001] of the Italian MIUR.

E. D. and H.-Y. W. contributed equally to this work.

# SUPPLEMENTAL MATERIAL

# Ultrafast adsorbate excitation probed with sub-ps resolution XAS

Elias Diesen,*,[1] Hsin-Yi Wang,[2] Simon Schreck,[2] Matthew Weston,[2] Hirohito Ogasawara,[3] Jerry LaRue,[4] Fivos Perakis,[2] Martina Dell'Angela,[5] Flavio Capotondi,[6] Luca Giannessi,[6] Emanuele Pedersoli,[6] Denys Naumenko,[6] Ivaylo Nikolov,[6] Lorenzo Raimondi,[6] Carlo Spezzani,[6] Martin Beye,[7] Filippo Cavalca,[2] Boyang Liu,[2] Jörgen Gladh,[2] Sergey Koroidov,[2] Piter S. Miedema,[7] Roberto Costantini,[5,8] Tony Heinz,[3,9] Frank Abild-Pedersen,[1] Johannes Voss,[1] Alan C. Luntz,[1] Anders Nilsson[2]

[1]SUNCAT Center for Interface Science and Catalysis, SLAC National Accelerator Laboratory, 2575 Sand Hill Road, Menlo Park, California 94025

[2]Department of Physics, AlbaNova University Center, Stockholm University, SE-10691 Stockholm, Sweden.

[3]SLAC National Accelerator Laboratory, 2575 Sand Hill Road, Menlo Park, California 94025

[4]Schmid College of Science and Technology, Chapman University, Orange, California 92866

[5]CNR-IOM, SS 14 – km 163.5, 34149 Basovizza, Trieste, Italy

[6]FERMI, Elettra-Sincrotrone Trieste, SS 14 – km 163.5, 34149 Basovizza, Trieste, Italy

[7]Deutsches Elektronen-Synchrotron DESY, Notkestrasse 85, 22607 Hamburg, Germany

[8]Physics Department, University of Trieste, Via Valerio 2, 34127 Trieste, Italy

[9]Department of Applied Physics, Stanford University, Stanford, California 94305

* E-mail: ediesen@stanford.edu

## A. DEFINITION OF TIME ZERO

Fig. S1 shows spectra at a few delays, each separated by 100 fs, around the onset of the pulse-induced effect. The spectral features remain similar to each other at the 1st, 2nd, and 3rd delays. At the 4th delay the intensity at the main peak (~ 288 eV) suddenly decreases with an increase of intensity on the lower energy region as indicated at ~287.5 eV. This suggests that time zero should be somewhere between 3rd and 4th delays. We therefore define the middle point between 3rd and 4th as the time zero, and name the 4th delay spectrum as collected between 0-100 fs, 5th delay spectrum as collected between 100-200 fs, and so on. The 1st to 3rd delay spectra are averaged together and further used as the unpumped delay spectrum.

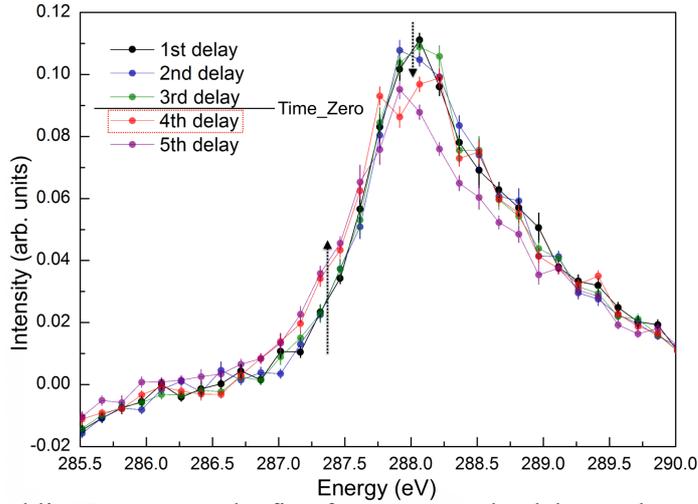

**Fig. S1**. Measured lin H spectra at the first few pump-probe delays, where the time gap between each delay is 100 fs.

## B. SPECTRA AT LONG TIME DELAYS

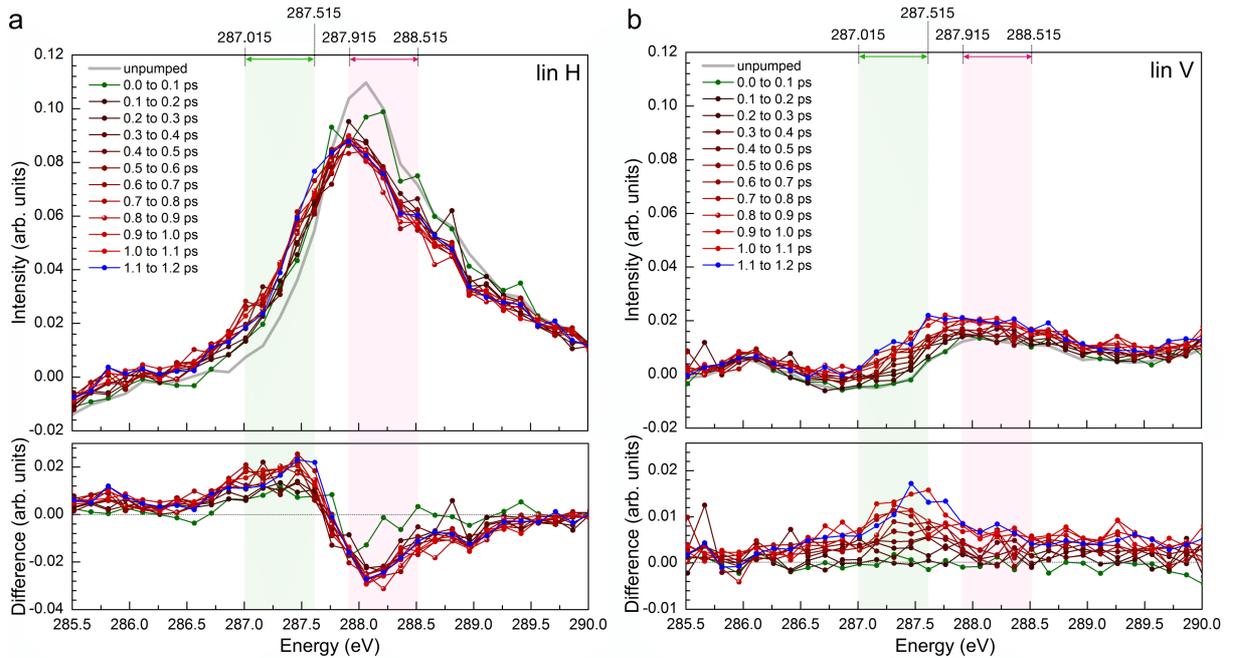

**Fig S2**. Measured (a) lin H spectra and (b) lin V spectra, averaged over the indicated pump-probe delays ranging from 0 to 1.2 ps. The lower panel shows the difference between each time delay spectrum and the unpumped spectrum.

## C. DENSITY FUNCTIONAL THEORY (DFT) METHOD

We use the DFT code Quantum ESPRESSO [1,2], which uses a plane-wave basis for the valence electrons and pseudopotentials to represent the core electrons. The RPBE functional was used due to its accuracy in determining surface adsorption energies [3]. All calculations are spin-unpolarized (restricted Kohn-Sham). We used a four-layer 2x2 slab model for initial geometry optimization, with 20 Å of vacuum separating the slab from its periodic image. The experimental lattice constant (2.7 Å) was used, and a 4x4x1 Monkhorst-Pack grid for Brillouin zone integration. For the excited-state calculations, we double the cell in the x- and y-directions to 4x4x4 (correspondingly reducing the k-point grid to 2x2x1) to reduce the interaction of the core hole with its periodic images. XAS spectra were computed using a refined 11x11x1 k-point grid, while keeping the density fixed (non-SCF/Harris calculation). Ultrasoft pseudopotentials [4] were used for atoms without core excitations. The plane-wave and density cutoffs were correspondingly chosen for the surface calculations to be 500 and 5000 eV, respectively. The geometries were optimized until all forces were less than 0.01 eV/Å, with the bottom two layers kept fixed.

The XAS is calculated using a Lanczos recursion Green's function technique, as implemented in the xspectra code [5–7]. For the spectrum calculations, a half-core-hole pseudopotential was used on the excited atom (transition-potential approach [8]) which has been shown to give reliable results [8,9]. For the core-excited C, norm-conserving half and full core-hole pseudopotentials were generated using the 'atomic' code included with the Quantum Espresso distribution [1]. Full and half-core pseudopotentials are generated by occupying the 1$s$ shell in the single-atom DFT solution by only, respectively, 1 and 1.5 electrons. The spectra were then shifted according to our calculated absorption onsets (Δ-Kohn-Sham method). All spectra were finally shifted by a common reference energy, chosen so as to match the experimental spectrum at negative delay with the calculated spectrum with the adsorbate in its vibrational ground state (see Fig. S4).

## D. CALCULATION OF VIBRATIONAL WAVEFUNCTIONS AND FRANCK-CONDON FACTORS

To accurately reproduce even the static ground state spectrum (at negative delay), we find that we need to consider both the spatial distribution of the initial vibrational state, and Franck-Condon transitions into different final internal stretch (IS) vibrational states. To this end we need the potential energy surfaces (PES) of the involved modes. The PES of the IS mode is calculated by varying the C-O distance while keeping the molecular center-of-mass fixed, and calculating the DFT total energy at a number of displacements. The IS PES is then fitted using a Morse potential $V(r) = D_e(1 - e^{-a(r-r_0)})$. The calculated potential is shown in Fig. S3. The vibrational states are then analytically given by:

$$\psi_n^{(gs)}(z) = N_n z^{\lambda - n - \frac{1}{2}} e^{-\frac{z}{2}} L_n^{(2\lambda - 2n - 1)}(z) \qquad (1)$$

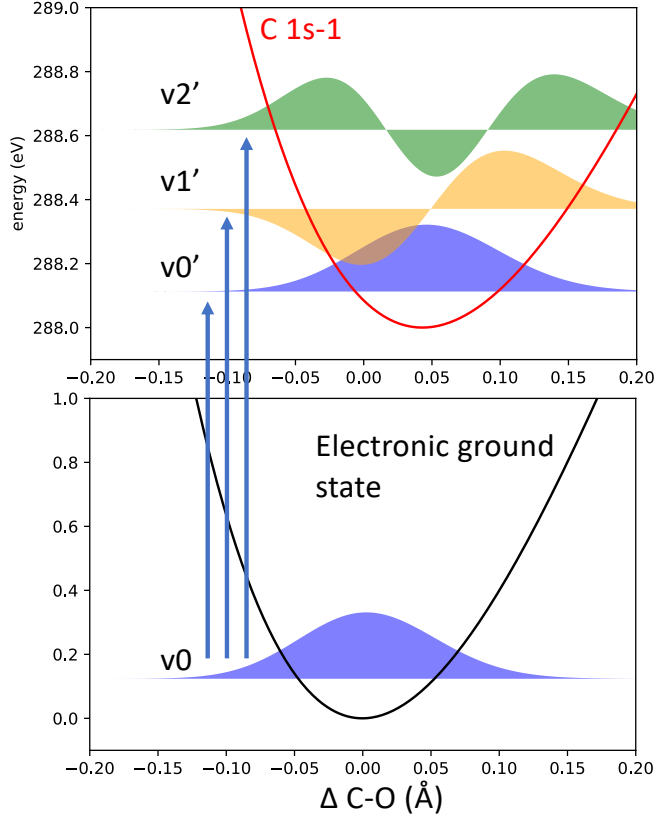

**Fig. S3**: The calculated internal stretch (IS) potential energy curves of the electronic ground state (black) and the C core-hole state (red), respectively. The lowest vibrational wavefunctions are also shown, with arrows illustrating Franck-Condon transitions to excited final states.

where $z = 2\lambda e^{-(x-x_0)}$; $x = ar$; $\lambda = \frac{\sqrt{2\mu D_e}}{a\hbar}$ ($\mu$ is the reduced mass of CO); $N_n = \sqrt{\frac{n!(2\lambda-2n-1)}{\Gamma(2\lambda-n)}}$ and $L_n^\alpha(z)$ is a generalized Laguerre polynomial.

To calculate the Franck-Condon factors, we also need the IS PES of the core-excited system. It was calculated in a similar way by varying the C-O distance, but with a full core-hole pseudopotential representing the C atom. It was also fitted with a Morse potential and is shown in Fig. S3. The vibrational states $\psi_n^{(exc)}$ are given by analogous expressions to Eq. 1, and we can calculate the IS Franck-Condon transition factors as $f_{ij} = \left|\left\langle \psi_j^{(exc)} \middle| \psi_i^{(gs)} \right\rangle\right|^2$

The frustrated rotation (FR) PES was assumed to be harmonic, using a ground-state vibrational analysis to obtain its fundamental frequency. We assume that the FR potential is cylindrically symmetric around the surface normal (z-axis), so that the motion can be described by two degenerate harmonic oscillators (one for rotation of angle $\theta_x$ around the x-axis and one $\theta_y$ around the y-axis). Then the potential can be written as a function of the polar angle $\theta$ alone: $V(\theta_x, \theta_y) = \frac{1}{2}I\omega(\theta_x^2 + \theta_y^2) = \frac{1}{2}I\omega\theta^2$ (here $I$ is the moment of inertia). We then get a 2D harmonic oscillator giving two-fold degenerate eigenstates, which can be conveniently written in terms of the angular momentum $M$ around the z-axis. The vibrational states are given by the expressions [10]:

$$\chi_{n,M}(\theta,\varphi) = C_{p,M}\theta^{|M|}e^{-\frac{\lambda\theta^2}{2}}L_p^{|M|}(\lambda\theta^2)e^{iM\varphi}$$

where $\lambda = I\omega/\hbar$, $p = (n - |M|)/2$, $C_{p,M} = \sqrt{\frac{2\lambda^{|M|+1}p!}{(p+|M|)!}}$ and $L_p^{|M|}$ is the generalized Laguerre function. $M$ takes the values $-n, -n + 2, ..., n - 2, n$ giving an $n + 1$-fold degeneracy of the n$^{th}$ level. The energy levels are given by $E_n = \hbar\omega(n + 1)$, with $n = 0,1,2, ....$

### E. CALCULATION OF SPECTRA FROM VIBRATIONAL STATES

Starting from the optimized structure, all coordinates **R** are kept fixed except the CO stretch, whose equilibrium distance we denote $r_0$. With the molecular center-of-mass fixed, we then calculate the XAS as described in section C, at different C-O distances $r$. The resulting elementary spectra, which we denote as $w(E, r)$, are shown in Fig. S4(a). We interpolate the result to get a continuous 2D function. For each vibrational state $n$, we then calculate the spectrum $w_n(E)$ by the following expression

$$w_n(E) = \sum_k f_{nk} \int w(E - (\epsilon_k - \epsilon_n), r)|\psi_n^{(gs)}(r)|^2 dr \qquad (2)$$

where $\psi_n^{(gs)}(r)$ is the vibrational wave function on the electronic ground state PES, $f_{nk}$ the Franck-Condon factor for the transition to the state $\psi_k^{(exc)}$, and $\epsilon_k, \epsilon_n$ the vibrational energies of the initial and final states. Fig. S4(b) shows the excellent agreement between the resulting spectrum $w_0(E)$ from the vibrational ground state with the experimental spectrum at negative pump-probe delay; the slightly higher experimental intensity around 288.8 eV is likely an effect of the higher experimental coverage (see Suppl. F below).

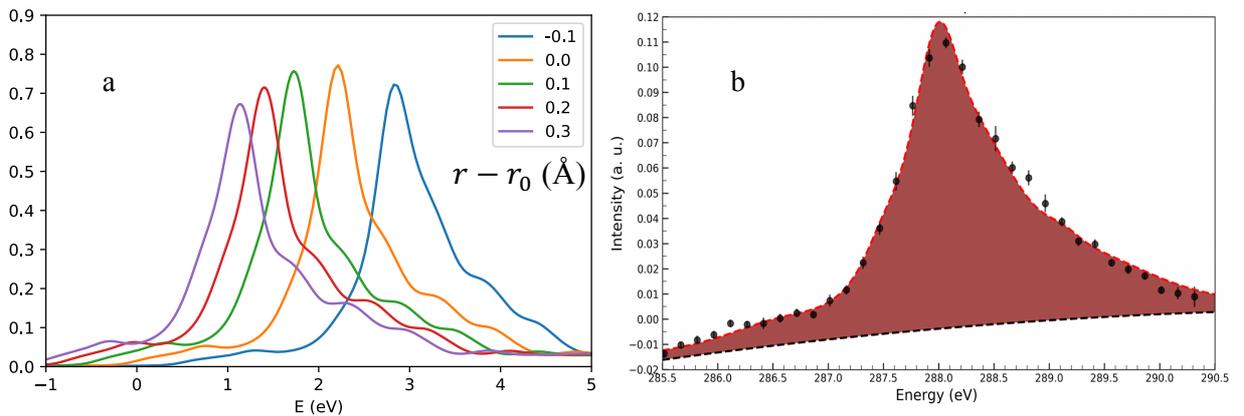

**Fig. S4:** (a) Calculated spectra for stretched C-O distances. (b) The experimental unpumped spectrum (dots) together with the calculated spectrum $w_0(E)$ (filled), which includes sampling of the spatial extension of the ground vibrational state and Franck-Condon transitions to excited vibrational states.

For the FR states, we similarly keep the molecular center-of-mass fixed, and calculate the XAS as a function of the angle $\theta$ to the surface normal. The elementary spectra $w(E, \theta_y)$ are shown in Fig. S5 for the three X-ray polarizations; $w(E, \theta_x)$ will be identical with the x- and y-polarizations interchanged. We approximate the dependence on the azimuthal angle $\varphi$ as $w(E, \theta, \varphi) \approx w(E, \theta_x)\cos\varphi + w(E, \theta_y)\sin\varphi$. For a single state $\chi_{n,M}$ we then get

$$w_{n,M}(E) = \int w(E, \theta, \varphi)|\chi_{n,M}(\theta, \varphi)|^2 \theta d\varphi d\theta \propto \int \left(w(E, \theta_x) + w(E, \theta_y)\right)|\chi_{n,M}(\theta)|^2 \theta d\theta$$

since $|\chi_{n,M}|^2$ is independent of $\varphi$. Writing $w(E, \theta) \equiv w(E, \theta_x) + w(E, \theta_y)$ we then calculate the total spectrum in a similar way to Eq. 2. We need to sum over all allowed $M$-values for each $n$, which yields the following expression

$$w_n(E) = \sum_M \sum_j f_{0j} \int w(E - \epsilon_j, \theta)|\chi_{n,M}(\theta)|^2 \theta d\theta$$

Since we do not consider a simultaneously excited IS state, we just need to consider Franck-Condon transitions $f_{0j}$ from the ground vibrational (IS) state $\psi_0^{(gs)}$.

To finally calculate the spectrum at an elevated temperature T, the spectra from the individual states are added together according to the Boltzmann distribution to give:

$$w(E, T) = \sum_n P_n(T) w_n(E) \quad \text{with} \quad P_n(T) = \frac{e^{-E_n/k_B T}}{\sum_n e^{-E_n/k_B T}}$$

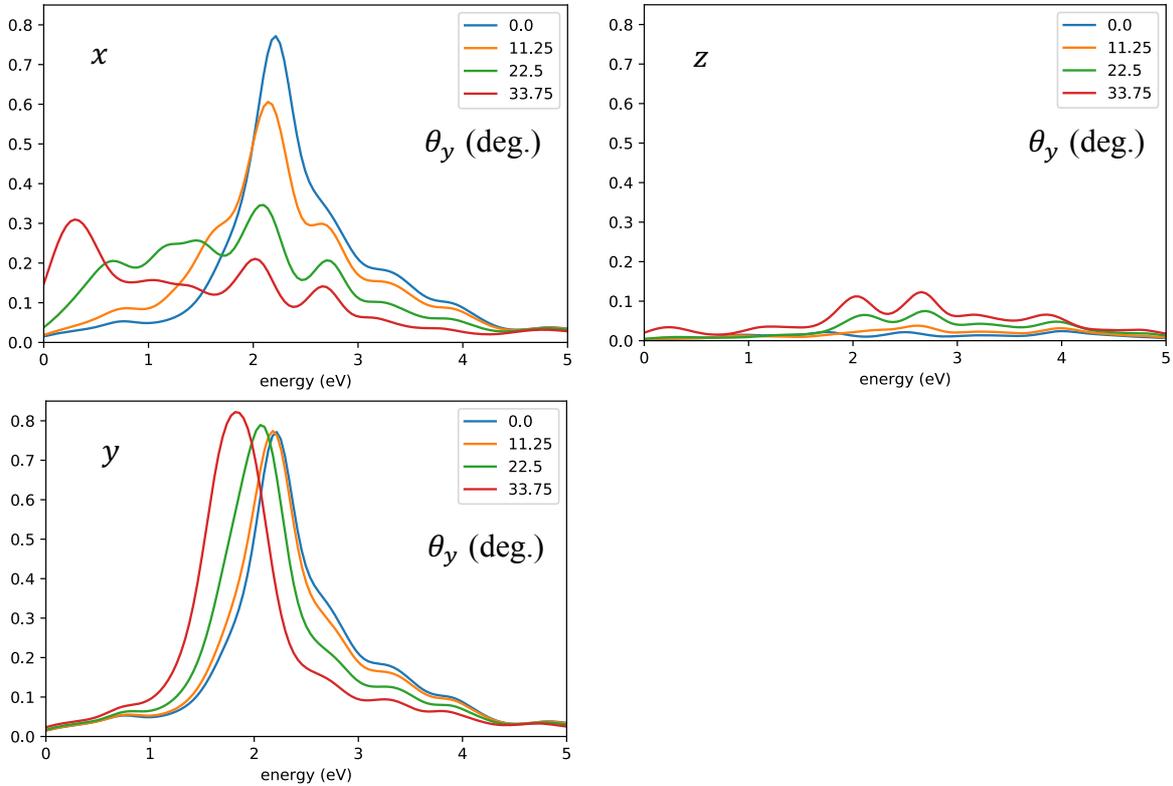

**Fig. S5:** Spectra for frustrated rotations for different X-ray polarizations. The molecule is rotated around the y-axis; the polarization of the X-ray beam is indicated.

## F. COVERAGE DEPENDENCE

To investigate if there is any significant difference between the model coverage of 0.25 ML and the experimental 0.66 ML, we perform certain calculations using a larger computational supercell where a 0.66 ML coverage can be constructed. This cell is shown in Fig. S6(b). There are two inequivalent CO in this structure, one occupying the center on-top site and six surrounding it in a tilted geometry. The two FR modes of the surrounding CO are no longer degenerate and have different frequencies from the central one. Fig. S6(a) shows the potentials for the different inequivalent frustrated rotations in this structure, showing that, while quantitatively slightly different from the low-coverage curve, the rotational motion will not be drastically altered. Calculating the ground state spectra from these molecules individually and then averaging the result gives the spectrum of S7, compared with the 0.25 ML case. (In these calculations, no initial state vibrational sampling or Franck-Condon factors were taken into account.) Fig. S7(a) further shows the spectra from different rotational angles from the optimized position, averaged over the seven molecules, all inequivalent rotations and the two parallel polarizations, compared to the low-coverage case. Fig. 7(b) shows the spectrum with polarization in the out-of-plane (lin V) direction.

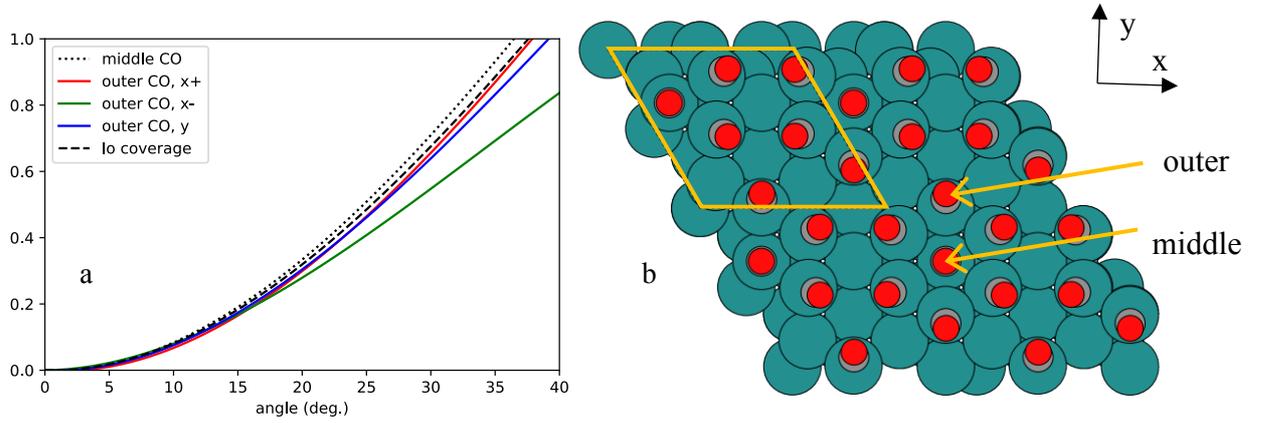

**Fig. S6:** (a) The potential energy curves for frustrated rotation at 0.66 ML coverage, compared to 0.25 ML (low coverage). There are three inequivalent rotations of the six outer CO molecules. Only the rotation around the x-axis in negative direction (meaning the molecule is rotating towards and past the upright position) has a noticeably different energy at high rotational angles than the low-coverage potential. (b) The high-coverage structure used for calculations. The unit cell is highlighted in yellow.

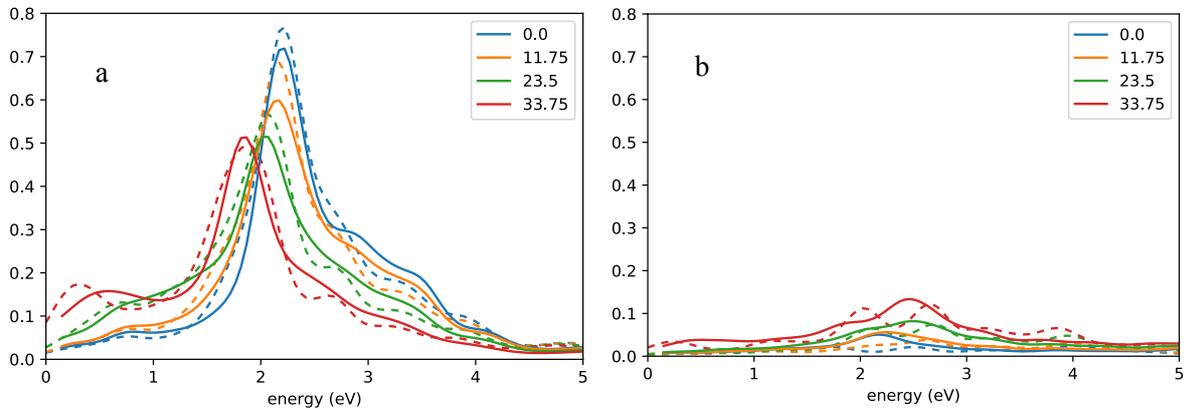

**Fig. S7:** Comparison of spectra for the high-coverage case (0.66 ML, full lines) with the low-coverage case (0.25 ML, dashed lines) for the FR mode at the indicated rotational angles. (a) Averaged over the two in-plane directions (lin H). (b) z-direction (lin V).

### G. TWO-TEMPERATURE MODEL

The common two-temperature (2T) model [11] describes the electrons and phonons as two coupled heat baths with depth- and time-dependent temperatures $T_e, T_{ph}$ whose time evolution is governed by the equations:

$$C_e \frac{\partial T_e}{\partial t} = \frac{\partial}{\partial z}\left(\kappa \frac{\partial T_e}{\partial z}\right) - g(T_e - T_{ph}) + S(z,t)$$

$$C_{ph} \frac{\partial T_{ph}}{\partial t} = g(T_e - T_{ph})$$

$$S(z,t) = I(t) e^{-z/\lambda}/\lambda$$

The electron heat capacity is calculated from the electron specific heat $\gamma$ as $C_e = \gamma T_e$. The heat capacity of the phonons is calculated using the Debye model:

$C_{ph} = 9nk_B \left(\frac{T_{ph}}{\Theta_D}\right)^3 \int_0^{\Theta_D/T_{ph}} \frac{x^4 e^x}{(e^x-1)^2}$, where $\Theta_D$ is the Debye temperature of the metal and $n$ the atomic density. The thermal conductivity is written as $\kappa = \kappa_0 \frac{T_e}{T_{ph}}$, with $\kappa_0$ taken as the low temperature value.

The laser pulse is modeled with a Gaussian time profile $I(t) = F\exp(-\frac{t}{2\sigma^2})/\sqrt{2\pi\sigma^2}$, where $F$ is the absorbed fluence on the sample and $\sigma = \Gamma/2.355$ with $\Gamma$ the full width at half maximum (FWHM). The initial temperature was set to 150 K. All other parameters are tabulated below (the Ru parameters are taken from [12]). The resulting temperatures in the surface layer are shown in Fig. S8. We have followed the traditional 2T model which only considers the phonon modes of the metal. Adding the phonon modes of the adsorbate CO can shorten the time for the $T_{ph}$ rise [13]. However, their inclusion does not affect the rise or peak in $T_e$.

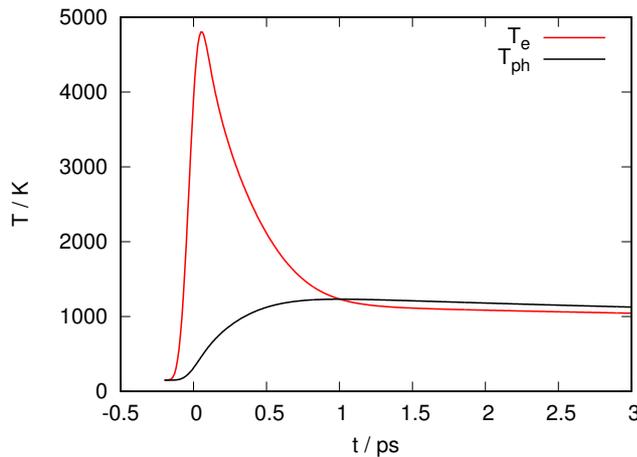

**Fig. S8:** Results from the 2T model; parameters are given in the text.

| | | |
|---|---|---|
| $\gamma$ | 400 Jm$^{-3}$K$^{-2}$ | Electron specific heat |
| $n$ | $7.4 \cdot 10^{28}$ m$^{-3}$ | Atomic density |

| $\Theta_D$ | 600 K | Debye temperature of Ru |
|---|---|---|
| $\kappa_0$ | 117 Wm$^{-1}$K$^{-1}$ | Electron heat conductivity |
| $g$ | 1.85· 10$^{18}$ Wm$^{-3}$K$^{-1}$ | El-ph coupling constant |
| $F$ | 91 Jm$^{-2}$ | Absorbed fluence |
| $\lambda$ | 6.9 nm | Optical penetration depth |
| $\Gamma$ | 100 fs | Pulse duration |

## H. AB-INITIO MOLECULAR DYNAMICS

To determine the spectrum from the equilibrated substrate-adsorbate system, which is formed after around 1 ps, we perform an *ab-initio* molecular dynamics (AIMD) simulation at a temperature of 1300 K. The dynamics was performed using the VASP [14] simulation package; the unit cell and all computational parameters were identical to the Quantum Espresso calculations. Starting at the equilibrium geometry with randomized initial velocities corresponding to a Maxwell-Boltzmann distribution at the desired temperature, we use a Langevin thermostat to simulate a thermal ensemble. After equilibrating the system for 1 ps, the dynamics was run for another 1 ps. Every 100 fs, the structure is saved and the XAS is calculated for that static structure using the Quantum Espresso package. These spectra are then averaged together for the final result.

The spectrum is shown in Fig. S9 for parallel (lin H) and perpendicular (lin V) polarization. The difference to the ground state spectrum is qualitatively similar to the experimental changes: a red-shift of the main peak in the lin H spectrum, and a growing intensity in the lin V. The effect is however more pronounced in the simulations. We attribute this mainly to the difference in coverage between the simulation and experiment. At this temperature CO is diffusing quite freely, and the presence of other adsorbates at the experimental 0.66 ML coverage will severely hinder this diffusion. It is also a known tendency of GGA-level DFT to overbind CO at hollow and bridge sites, so that the relative population of the different sites is likely not quantitatively accurate in the AIMD, even at a coverage of 0.25 ML. This would lead to a more drastic spectrum changes than seen experimentally. We do however find that the AIMD spectrum shows the correct qualitative features, and indicates that the spectral changes evolving over a ps timescale is consistent with general thermal dynamics of the substrate-adsorbate system.

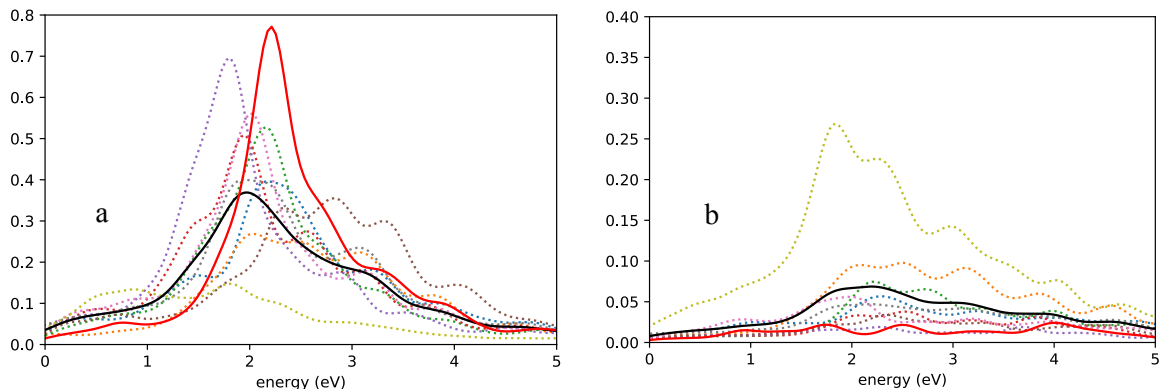

**Fig. S9:** Results of the AIMD simulated T=1300 K spectrum (black) compared to the ground state spectrum (red). The individual contributions to the simulated high-temperature spectrum are shown in dotted lines, illustrating the wide variety of bonding situations the CO molecule explores during 1 ps.

## I. ADSORBATE HEATING: MASTER EQUATION MODEL

In an electronic friction model, the adsorbate is assumed to interact with excited electron-hole pairs in a thermal distribution in the substrate that couple to the relevant adsorbate mode, with motion taking place on a single potential energy surface (electronic ground state). A simple estimate of the energy transfer into a single mode is given by a classical master equation approach [15] describing the internal energy $U_v(t)$ of a harmonic oscillator coupled to a heat bath at a prescribed (time-dependent) temperature $T_e(t)$. The temporal evolution of $U_v(t)$ is then given as

$$\frac{dU_v(t)}{dt} = -\eta(U_v(t) - U_e(t))$$

where $U_e(t) = \hbar\omega(\exp\left[\frac{\hbar\omega}{k_B T_e(t)}\right] - 1)^{-1}$. This equation is solved numerically with $T_e(t)$ taken from the 2T model and a few different choices for the coupling parameter $\eta$. An adsorbate temperature is defined by $U_v(t) = \hbar\omega(\exp\left[\frac{\hbar\omega}{k_B T_v(t)}\right] - 1)^{-1}$.

The results are shown in Fig. S10. It is seen that a very strong frictional coupling of $\eta \approx 30$ cm$^{-1}$ or ~ 900 GHz is needed to get a high adsorbate temperature within a few hundred fs, corresponding to a relaxation time constant $\tau = \frac{\hbar}{\eta} \approx 170$ fs. By comparison, results in section K gives a relaxation time constant of 2- 4 ps for CO on Ru.

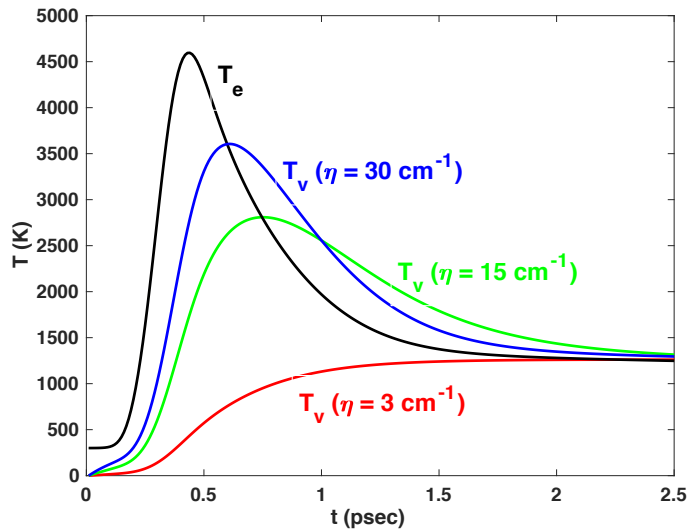

**Fig. S10:** Results of the master equation model. To get a significant $T_v$ within a few hundred fs, a frictional coupling of $\eta = 20 - 30$ cm$^{-1}$ is necessary.

## J. ADSORBATE HEATING: ION RESONANCE MODEL

A common picture for the coupling between excited electrons and adsorbate motion is the ion resonance model, where a single energetic electron resonantly populate an unoccupied orbital of the adsorbate. This mechanism is behind the DIET and DIMET desorption processes [16]. Within this picture we can get an estimate of the energy transfer to the adsorbate IS mode, due to resonant population of the adsorbate LUMO ($2\pi^*$) by scattering of excited electrons from the substrate into the resonance. To this end, we calculate the PES $V_1$ for the CO using the linear-expansion ΔSCF method [17], which ensures that one $2\pi^*$-orbital is populated by one electron. The GPAW code [18,19] was used for the excited-state PES calculation, using identical parameters to the Quantum Espresso calculations in the rest of this work. The resulting PES, fitted with a Morse potential, is shown in Fig. S11(a).

We treat the dynamics as occurring on either the ground state ($V_0$) or the excited state ($V_1$) PES, with discrete jumps between the two electronic states. The lifetime $\tau_1$ of the resonance is not known in detail, but is estimated to be in the range of a few fs. The time $\tau_0$ spent in the ground state between two excitations is also not known. We therefore study a range of possible lifetimes. In the real system, the lifetime and excitation rate would be strongly dependent on the electron temperature and thus time-dependent. To simplify our model, we consider only fixed $\tau_1$ and $\tau_0$, and determine how much energy the adsorbate obtains during a fixed time of $t_{max} = 200$ fs, roughly the width of the peak in electronic temperature.

We treat the dynamics on each PES quantum-mechanically. To this end we write the time-evolution operator for dynamics on a single PES as $U_n(t', t'') = \exp(-iH_n(t'' - t')/\hbar)$, where $H_n = T + V_n$ is the Hamiltonian for evolution on the ground ($n = 0$) or excited ($n = 1$) electronic state. For a number of jumps at times $t_1, t_2, t_3 \dots, t_k < t_{max}$, chosen from an exponential distribution with width according to the chosen lifetimes $\tau_1$ and $\tau_0$, we then write the system dynamics as

$$|\Psi_{final}\rangle = U_n(t_k, t_{max}) \dots U_1(t_3, t_4) U_0(t_2, t_3) U_1(t_1, t_2) U_0(t_0, t_1) |\Psi_0\rangle.$$

Since the eigenstates of the Morse potential is analytically known, numerical implementation of this expression is straightforward. Denoting the eigenstates of $H_0$ as $\psi_k$ and of $H_1$ as $\varphi_l$, we can write $U_0(t', t'') = \sum_k |\psi_k\rangle \exp(-\frac{i\omega_{0,k}(t''-t')}{\hbar})\langle\psi_k|$ and $U_1(t', t'') = \sum_l |\varphi_l\rangle \exp(-\frac{i\omega_{1,l}(t''-t')}{\hbar})\langle\varphi_l|$, so that e.g. one single excitation, propagation on the excited PES, and deexcitation, is given by

$$|\Psi_{final}\rangle = U_1(t_1, t_2) U_0(t_0, t_1) |\Psi_0\rangle = \sum_{k,l,k'} |\psi_{k'}\rangle \sqrt{f_{k'l}} \sqrt{f_{kl}} \exp(-\frac{i\omega_{1,l}(t_2-t_1)}{\hbar}) \exp(-\frac{i\omega_{0,k}(t_1-t_0)}{\hbar}) \langle\psi_k|\Psi_0\rangle$$

where $f_{kl}$ are the Franck-Condon factors for the transition: $f_{kl} = |\langle\varphi_l|\psi_k\rangle^2|$.

The expectation value of the final energy, given for each run by $\langle E_{final}\rangle = \langle \Psi_{final}|H_0|\Psi_{final}\rangle$ and averaged over 500 realizations, is shown in Fig. S11b as a function of the excited state lifetime $\tau_1$, for different values of the time $\tau_0$ between the excitations. In order to get appreciable energy into the IS, we need both unrealistically large $\tau_1$, on the order of tens of fs, and very short $\tau_0$. Given the coarse nature of this model, the results should only be seen as an order of magnitude estimate. However, given that realistic $\tau_1$ is on the order of a few fs, and $\tau_0$ should be much larger than $\tau_1$ under normal conditions (see e.g. section 6.2 of Ref. [15]), we conclude that this mechanism is unable to explain the observed rapid adsorbate excitation, notwithstanding that the high energy of the LUMO makes resonant population of it unlikely.

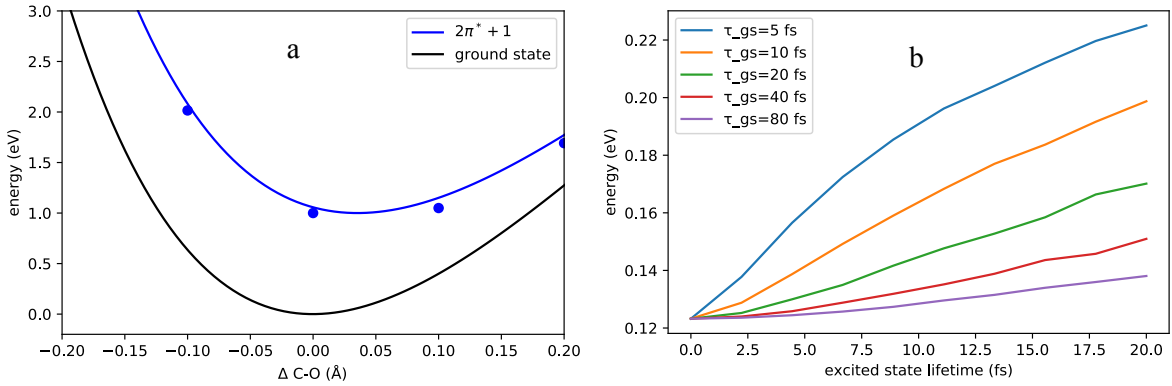

**Fig. S11:** (a) PES for the resonant $2\pi^*+1$ state ($V_1$) and the electronic ground state ($V_0$). The dots are the calculated linear-expansion ΔSCF values, the full line is a Morse potential fit. (b) Final energy in the vibration after 200 fs.

### K. CALCULATION OF LIFETIMES OF ADSORBATE MODES

We calculate the theoretical lifetimes of the CO/Ru(0001) adsorbate vibrational modes using the DFPT [20] method implemented in Quantum Espresso. Details on the method and its implementation can be found in e.g. [21,22]. The phonon linewidth $\gamma_{q\nu}$ is calculated by evaluating the Fermi-golden-rule-like expression

$$\gamma_{q\nu} = 2\pi\omega_{q\nu}\sum_{ij}\int\frac{d^3\mathbf{k}}{\Omega_{BZ}}|g_{q\nu}(\mathbf{k},i,j)|^2\delta(e_{\mathbf{q},i}-e_F)\delta(e_{\mathbf{k+q},j}-e_F) \quad (3)$$

where $g_{q\nu}(\mathbf{k},i,j)$ are the electron-phonon coefficients

$$|g_{q\nu}(\mathbf{k},i,j)| = \left(\frac{\hbar}{2M\omega_{q\nu}}\right)^{1/2}\left\langle\psi_{i,\mathbf{k}}\left|\frac{dV_{SCF}}{du_{q\nu}}\cdot\epsilon_{q\nu}\right|\psi_{j,\mathbf{k+q}}\right\rangle$$

describing the coupling between the electronic bands $i,j$ due to the variation of the potential $V_{SCF}$ as the ions are vibrating. $u_{q\nu}$ are ionic displacements along the eigenvector $\epsilon_{q\nu}$ of the dynamical matrix corresponding to a phonon $\nu$ with frequency ω and wavevector **q**. In contrast to many studies of adsorbate vibrational lifetime that were carried out using IR excitation and hence only sensitive to the behavior at $\boldsymbol{q} = 0$, the lack of a dipole selection rule for electron-hole pair scattering means excitation can happen at any $\boldsymbol{q}$. We thus calculate $\gamma_{q\nu}$ for a grid of points in the first Brillouin zone and average the result:

$$\bar{\gamma}_\nu = \int \frac{d^2\boldsymbol{q}}{\Omega_{BZ}} \gamma_{q\nu}$$

from which we get the lifetime as $\tau_\nu = \hbar/\bar{\gamma}_\nu$. The double-delta expression in Eq. 3 is evaluated by replacing the delta functions by Gaussian functions, and different widths σ are used in order to study convergence. To ensure a good enough sampling of the electronic structure for reliable coupling coefficients, the slab was extended to six layers depth, and a k-point grid of 48x48x1 points was used for evaluating the linewidth using the interpolation scheme described in [22]. Convergence was checked by increasing the grid density to 72x72x1, which had negligible effect on the results.

Fig. S12 shows the resulting linewidths in cm$^{-1}$ for the IS, SA and the two FR modes, calculated for a few different high-symmetry q-points, as functions of the broadening parameter. The results are quite independent of broadening σ down to around 0.2 eV, while being numerically unstable for the smallest broadenings. For the value of the lifetimes reported in Table S1, we choose σ=0.3 eV, and take the average of the two FR modes. We have confirmed that our calculation method gives similar results for CO/Cu(100) as in [23].

For CO/Cu(100) a significant temperature dependence of calculated IS lifetimes has been reported [23]. Temperature effects can be accounted for in an approximate way by identifying the broadening used in the δ-functions with a temperature [24]. For low temperatures this can be inadequate since the temperature broadening is not Gaussian but given by Fermi-Dirac distributions; furthermore the high frequency of the IS mode means it cannot safely be neglected, and a more complete expression than Eq. 3 should be used (e.g. Eq. 4 of [21]). Our results at σ=0.3 eV corresponds to T≈3000 K where these issues are of less concern. Furthermore, changing the broadening has little effect on the linewidth in Fig. S12, which means the temperature dependence of the $\gamma_{q\nu}$ would also be weak, even if the quantitatively more exact thermal broadening were used to determine $\gamma_{q\nu}$. Physically, the lack of a temperature effect can easily be rationalized by the featureless CO PDOS in the relevant energy region around the Fermi level, in contrast to the CO/Cu(100) case where the $2\pi^*$ resonance is only around 2 eV above $E_F$.

A more detailed analysis of adsorbate vibrations has been performed recently for CO/Cu(100) [25], which removes a well-known divergence at the Γ-point inherent in the double-delta approximation [21]. Numerically we find this divergence unproblematic down to at least 0.2 eV broadening, and the linewidth is only weakly dependent on $\boldsymbol{q}$, as expected for modes dominated by adsorbate vibration at low coverage.

We perform the same calculation for the case of 1 ML coverage, using a 1x1x6 slab and doubling the k-point grid. It is notable that the linewidths $\gamma_{q\nu}$ vary significantly more over the Brillouin zone for 1 ML coverage, especially for the IS mode where the coupling at the Γ-point is quite weak, in line with the result of [25]; the averaged quantity $\bar{\gamma}_\nu$ is much less affected. Overall, the 1 ML case gives somewhat shorter lifetimes than 0.25 ML, however the change is within a factor of 1.5. We expect that the lifetimes in the experimental 0.67 ML system will fall in between the 0.25 and 1 ML case.

| Coverage | IS | SA | FR |
|---|---|---|---|
| 0.25 ML | 2.8 | 6.2 | 4.2 |
| 1 ML | 2.2 | 4.2 | 3.5 |

**Table S1:** Calculated lifetimes $\tau$ (ps) of relevant CO/Ru(0001) adsorbate modes

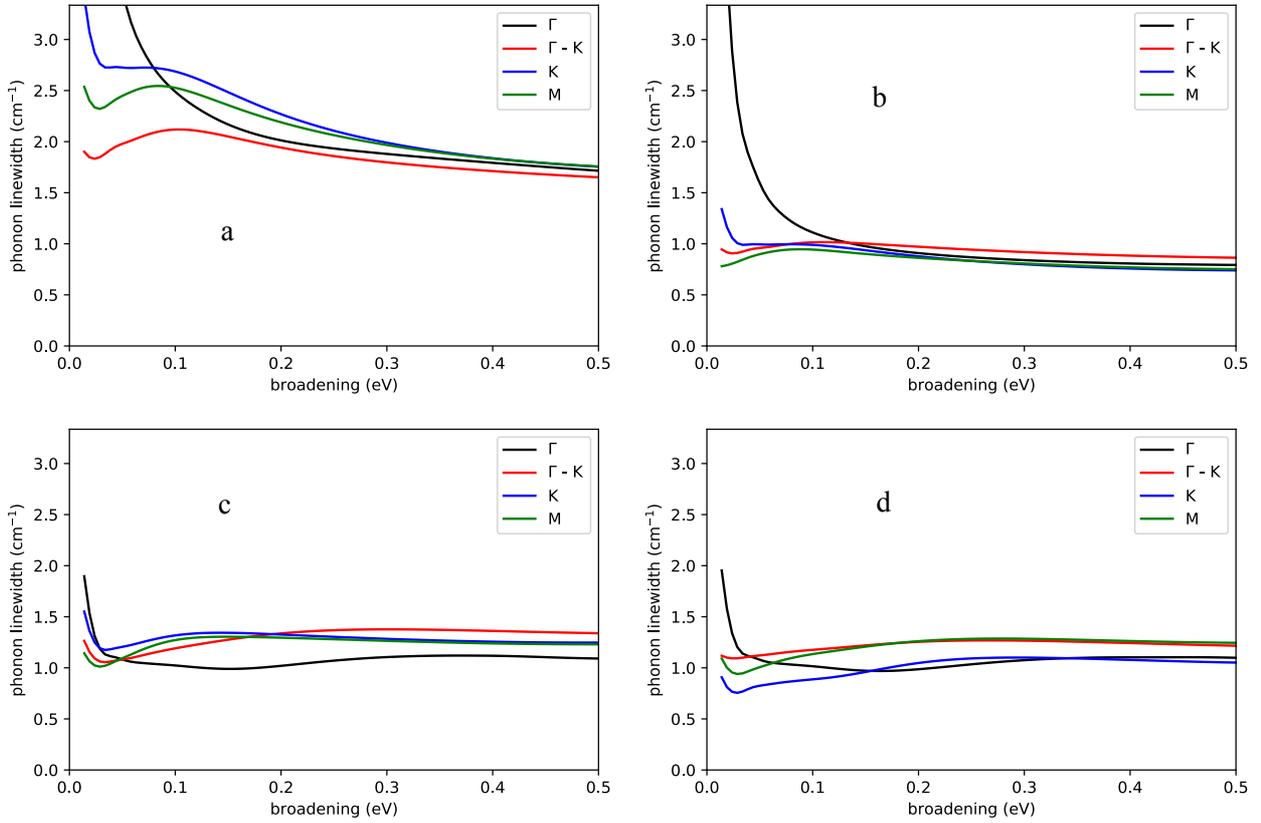

**Fig. S12:** q-point resolved phonon linewidths $\gamma_{q\nu}$ for CO/Ru(0001) at 0.25 ML coverage, as functions of the Gaussian broadening of the delta functions. (a) Internal stretch (IS) mode, (b) surface-adsorbate (SA) mode, (c) and (d) the two frustrated rotational (FR) modes.

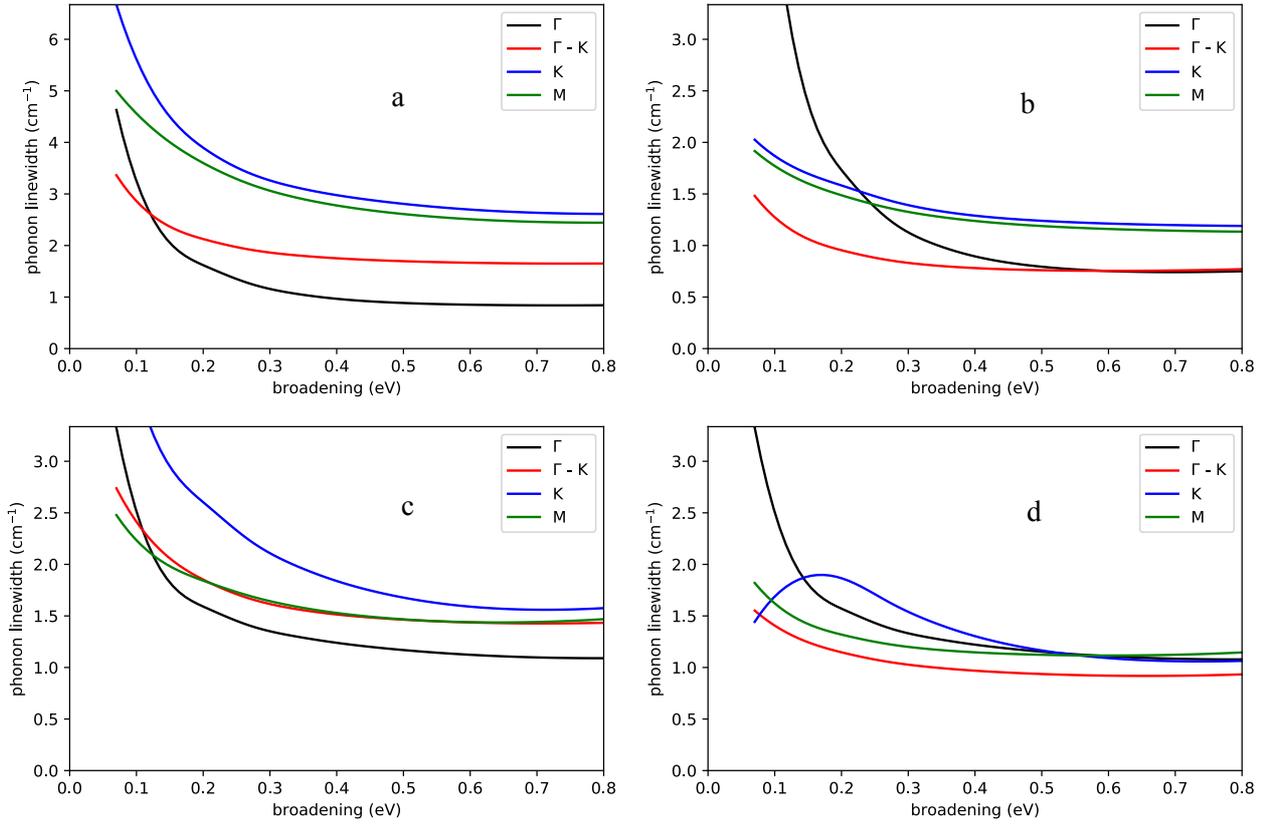

**Fig. S13:** Same as Fig. S12 for 1 ML coverage